\documentclass[aps,prd,twocolumn]{revtex4}

\begin{document}

\title{Thermal self-energies using light-front quantization}

\author{H. Arthur Weldon}

\affiliation{Department of Physics, West Virginia University,
Morgantown, West Virginia 26506-6315}

\date{\today}

\begin{abstract}
A recent paper by Alves, Das, and Perez contains a calculation of the one-loop
self-energy in $\phi^{3}$ field theory at $T\neq 0$ using light-front quantization and
concludes that the  self-energy is different than the conventional answer and is not
rotationally invariant. The changes of variable displayed below show that
despite the complicated appearance of the thermal self-energy in light-front
variables, it is exactly the same as the conventional result.  
\end{abstract}

\pacs{11.10.Wx, 11.10.Kk, 12.38.Lg}

\maketitle

In Ref. \cite{ADP}, Alves, Das, and Perez  introduce the technique of
light-front quantization into thermal field theory  using a heat bath at rest. 
As shown in
\cite{AW} the appropriate quantization 
   evolves the system in the
$x^{+}=(x^{0}\!+\!x^{3})/\sqrt{2}$ coordinate while keeping constant $x^{1}, x^{2},$ and
$x^{3}$. (Normal light-front quantization keeps $x^{1}, x^{2},$
and $x^{-}=(x^{0}\!-\!x^{3})/\sqrt{2}$ constant.)   The  momenta conjugate to  $x^{+},
x^{1}, x^{2}, x^{3}$ are $k^{0}, k^{1}, k^{2}, k^{+}$ as can be seen from the identity
\begin{displaymath}
k^{0}x^{0}\!-k^{3}x^{3}\!-{\bf k}_{\bot}\cdot{\bf
x}_{\bot}=\sqrt{2}\,k^{0}x^{+}\!-\sqrt{2}\,k^{+}x^{3} \!-{\bf k}_{\bot}\cdot{\bf x}_{\bot}.
\end{displaymath}
In the imaginary time propagator, $x^{+}$ is made negative imaginary: $-i\beta\le
\sqrt{2}x^{+}\le 0$.
 In the Fourier transform of the  propagator 
$k^{0}=i2\pi nT$ whereas
$k^{+}$ and ${\bf k}_{\bot}$ are real. The relation
\begin{displaymath}
(k^{0})^{2}-(k^{3})^{2}-k_{\bot}^{2}=2\sqrt{2}k^{0}k^{+}-2(k^{+})^{2}-k_{\bot}^{2}
\end{displaymath}
immediately leads to the  propagator
\begin{displaymath}
G(k^{+}, k_{\bot},n)={1\over i4\sqrt{2}\pi n k^{+}-2(k^{+})^{2}-\omega_{k}^{2}},
\end{displaymath}
where $\omega_{k}^{2}=k_{\bot}^{2}+m^{2}$ is the transverse energy. 

One of the
interesting calculations performed by Alves, Das, and Perez \cite{ADP} using this
propagator is the one-loop   self-energy for a scalar field theory with
self-interaction
$g\phi^{3}/3!$.  The result of  performing the summation over the loop integer $n$ is
given in  Eq. (40) of Ref. \cite{ADP}:
\begin{equation}
\Pi(p)\!=\!{g^{2}\over 8}\!\!\int \!{dk^{+}d^{2}k_{\bot}\over(2\pi)^{3}}
\,{\coth(X_{1}/2T)-\coth(X_{2}/2T)\over Y}.
\end{equation}
During the summation the external variable $p^{0}$ is an integer multiple 
of $2\pi iT$, but after the summation $p^{0}$ is continued to real values. 
The quantities $X_{1}, X_{2},$ and $Y$ are complicated functions of the
integration variables $k^{+}, {\bf k}_{\bot}$ and of the external variables $p^{0},
p^{+}, {\bf p}_{\bot}$:
\begin{eqnarray}
X_{1}=&&{\omega_{k}^{2}+2(k^{+})^{2}\over 2\sqrt{2}k^{+}}\nonumber\\
X_{2}=&&{\omega_{k+p}^{2}+2(k^{+}\!+p^{+})^{2}\over
2\sqrt{2}(k^{+}+p^{+})}\nonumber\\
Y=&&2\sqrt{2}\,k^{+}(k^{+}+p^{+})[-X_{1}+X_{2}-p^{0}].\nonumber
\end{eqnarray}
The self-energy (1) looks quite different from the usual result and is not manifestly
invariant under  $O(3)$ rotations of the external
momentum ${\bf p}$.

The following will describe a change of integration variable from 
$k^{+}$ to a new variable $k^{3}$ that is chosen to make the self-energy 
 a function of the two variables
$p^{0}$ and ${\bf p}^{2}=p_{\bot}^{2}+(\sqrt{2}p^{+}-p^{0})^{2}$.  
The final answer is the sum of Eqs. (2), (3), (4), and
(5).

(1a) For the term $\cosh(X_{1}/2T)/Y$ in Eq. (1),  when  $k^{+}$
is positive  change to a new integration variable $k^{3}$ defined by
\begin{displaymath}
k^{+}={1\over\sqrt{2}}\bigg[k^{3}+\sqrt{m^{2}+k_{\bot}^{2}+(k^{3})^{2}}\bigg].
\end{displaymath}
The range of $k^{3}$ is $-\infty\le k^{3}\le\infty$. The Jacobian of the
transformation is $dk^{+}/dk^{3}=k^{+}/E_{k}$,  where
$E_{k}=\sqrt{m^{2}+k^{2}}$ is the square root displayed above.  Under this change, 
\begin{displaymath}
X_{1}=E_{k};\hskip0.6cm 
Y=k^{+}\big[(E_{k+p})^{2}-(p^{0}+E_{k})^{2}\big],
\end{displaymath}
where
$E_{k+p}=\sqrt{m^{2}+({\bf k}+{\bf p})^{2}}$. The factor $k^{+}$ in the Jacobian
cancels a similar factor in $Y$ and yields
a contribution to the self-energy
\begin{equation}
 \Pi_{1a}={g^{2}\over 8}\!\!\int {d^{2}k_{\bot}\over (2\pi)^{3}}\!
\int_{-\infty}^{\infty}\!{dk^{3}\over E_{k} }\;{\coth(E_{k}/2T)\over
(E_{k+p})^{2}-(p^{0}\!+\!E_{k})^{2}}.
\end{equation}
This integrand is invariant
invariant under simultaneous rotations of the vectors ${\bf k}$ and ${\bf p}$.
Thus $\Pi_{1a}$ depends only on $|{\bf p}|$ and $p^{0}$.  

(1b) For  term $\cosh(X_{1}/2T)/Y $in Eq. (1), when $k^{+}$ is negative
make the change of variable
\begin{displaymath}
k^{+}={1\over\sqrt{2}}\bigg[k^{3}-\sqrt{m^{2}+k_{\bot}^{2}+(k^{3})^{2}}\bigg],
\end{displaymath}
where  $-\infty\le k^{3}\le\infty$.
The Jacobian of the transformation is
$dk^{+}/ dk^{3}=-k^{+}/E_{k}$ and 
\begin{displaymath}
X_{1}=-E_{k};\hskip0.6cm 
Y=k^{+}\big[(E_{k+p})^{2}-(p^{0}-E_{k})^{2}\big].
\end{displaymath}
The corresponding self-energy contribution is
\begin{equation}
\Pi_{1b}={g^{2}\over 8}\int {d^{2}k_{\bot}\over (2\pi)^{3}}
\int_{-\infty}^{\infty}{dk^{3}\over E_{k} }\;{\coth(E_{k}/2T)\over
(E_{k+p})^{2}-(p^{0}\!\!-\!E_{k})^{2}}.
\end{equation}
The sum of Eqs. (2) and (3) is an even function of $p^{0}$.

(2a) In the second term in Eq. (1),  $\cosh(X_{2}/2T)/Y$, when 
$k^{+}\!+\!p^{+}>0$ then change to $k^{3}$ given by
\begin{displaymath}
k^{+}={1\over\sqrt{2}}\bigg[k^{3}\!-\!p^{0}+\sqrt{m^{2}+({\bf k}_{\bot}\!+\!{\bf
p}_{\bot})^{2} +(k^{3}\!+\!p^{3})^{2}}\bigg],
\end{displaymath}
where $-\infty\le k^{3}\le \infty$. Using 
$dk^{+}/dk^{3}=(k^{+}\!\!+\!p^{+})/ E_{k+p}$, and
\begin{displaymath}
X_{2}=E_{k+p};\hskip0.6cm Y=(k^{+}\!+p^{+})\big[(p^{0}\!-E_{k+p})^{2}\!-E_{k}^{2}\big].
\end{displaymath}
This contribution is
\begin{equation}
\Pi_{2a}={g^{2}\over 8}\!\int {d^{2}k_{\bot}\over (2\pi)^{3}}
\!\int_{-\infty}^{\infty}{dk^{3}\over E_{k+p} }\;{\coth(E_{k+p}/2T)
\over E_{k}^{2}-(p^{0}\!-E_{k+p})^{2}}.
\end{equation}

(2b) In the second term in Eq. (1), $\cosh(X_{2}/2T)/Y$, if $k^{+}+p^{+}<0$ make the change
of variable
\begin{displaymath}
k^{+}={1\over\sqrt{2}}\bigg[k^{3}\!-\!p^{0}-\sqrt{m^{2}+({\bf k}_{\bot}\!+\!{\bf
p}_{\bot})^{2} +(k^{3}\!+\!p^{3})^{2}}\bigg],
\end{displaymath}
where $-\infty\!\le \!k^{3}\!\le\! \infty$. Since
$dk^{+}/dk^{3}=-(k^{+}\!+p^{+})/E_{k+p}$, and
\begin{displaymath}
X_{2}=-E_{k+p},\hskip0.5cm Y=(k^{+}\!+p^{+})\big[(p^{0}+E_{k+p})^{2}-E_{k}^{2}\big],
\end{displaymath}
the contribution to the self-energy is
\begin{equation}
\Pi_{2b}={g^{2}\over 8}\!\int\! {d^{2}k_{\bot}\over (2\pi)^{3}}\!
\int_{-\infty}^{\infty}{dk^{3}\over E_{k+p}}\;{\coth(E_{k+p}/2T)
\over E_{k}^{2}-(p^{0}+E_{k+p})^{2}}.
\end{equation}
The sum of Eq. (4) and (5) is an even function of $p^{0}$. 

The sum of these four contributions Eq. (2), (3), (4), and (5)
is the standard answer for the self-energy \cite{Das}.
Therefore the light-front formulation is a different, and in some cases a more
efficient \cite{AW}, way of organizing the calculation, but the results are the same.

\begin{acknowledgments}
This work was supported in part by  the U.S. National Science
Foundation under Grant No. PHY-0099380.
\end{acknowledgments}

\end{document}